\documentclass[12pt,preprint]{aastex}

\begin{document}

\title{Possible High-Energy Neutrinos from the Cosmic Accelerator \\
RX J1713.7-3946}

\author{J. Alvarez-Mu\~niz}
\affil{Bartol Research Institute,\\ University of Delaware,
Newark, DE 19716, USA.}

\and

\author{F. Halzen}
\affil{University of Wisconsin, Department of Physics,\\ 
1150 University Avenue, Madison, WI 53706, USA.}

\begin{abstract}
The observation of TeV-gamma rays of $\pi^0$ origin from the supernova 
remnant RX~J1713.7-3946 might have revealed the first specific site where 
protons are accelerated to energies typical of the main component of the 
cosmic rays. In this letter we calculate the high-energy neutrino flux 
associated with this source to be 40 muon-type neutrinos per 
kilometer-squared per year. We perform the same calculations for other known 
sources of TeV-gamma rays and show how neutrino observations can 
establish whether the TeV-gamma rays emitted by blazars and supernova 
remnants are the decay products of neutral pions and thus unequivocally 
establish the sources as cosmic accelerators.
\end{abstract}

\keywords{acceleration of particles --- cosmic rays ---
gamma rays: theory --- neutrinos --- supernova remnants}

\section{Introduction}

Supernova remnants (SNRs), 
such as the Crab \citep{WeekesCrab},
Cassiopeia A \citep{HEGRACasA}
and blazars, such as Mkn 421 \citep{PunchMkn421} and Mkn 501
\citep{QuinnMkn501},
emit gamma rays with energies up to tens of TeV, (see \citep{ong}, 
\citep{catanese}, and \citep{Pohl} for reviews). 
Unfortunately, these
observations cannot by themselves establish whether these high-energy 
photons are produced by electron beams via inverse Compton
scattering \citep{Pohl}, or by
proton beams via the production and subsequent decay
of neutral pions \citep{Physrep,Pohl}. The
mechanisms for the production of the highest energy photons by purely
electromagnetic processes do not naturally accommodate TeV-photons but 
can be stretched to explain the observed spectrum. 
The question is central to the problem of the origin of cosmic
rays and it has been pointed out that neutrino observations can
unequivocally settle this debate \citep{Physrep}. In hadronic mechanisms 
the high-energy photons are accompanied by neutrinos because 
charged pions are produced along with $\pi^0$'s.

Recently the CANGAROO collaboration has produced 
evidence for the observation of $\gamma$-rays of $\pi^0$ origin from the 
SNR RX J1713.7-3946 \citep{CANGAROO-RX}. We here show 
how a lower limit on the 
neutrino flux from this source can be established from energy 
conservation and routine particle physics. We show that this limit is 
accessible by a northern hemisphere neutrino observatory 
such as Antares \citep{Antares} operated over several years. Obviously, if 
photons are partially absorbed in the source or on the interstellar 
medium, this calculation underestimates the actual neutrino flux.

The true nature of RX J1713.7-3946 is still a matter of discussion
\citep{Butt01}. Our calculation illustrates the possibility to disentangle 
electromagnetic and hadronic sources of high-energy radiation 
observationally. We will establish, in a largely model-independent way, 
a direct relation between the lower limit on the neutrino flux and the 
TeV-gamma ray observations for sources other than RX J1713.7-3946. We 
compute the response of a kilometer-scale neutrino observatory such as 
IceCube \citep{ICECUBE} to that flux, and conclude that the event 
rates are adequate to confirm or rule out the hadronic origin of the
highest energy photons. Although the flux from a single source may still 
be small, this conclusion is credible because a neutrino telescope will be
operated for a decade.

We should emphasize that, even if 
RX J1713.7-3946 is a truly cosmic ray accelerator, it is likely that 
the dominant source of the cosmic rays has not been 
identified. The higher energies might be associated with older supernova 
remnants that may be astronomically uninteresting and have escaped 
attention. Because of their restricted angular resolution atmospheric 
Cherenkov telescopes have scrutinized a limited number of sources. 
Moreover, TeV-gamma ray instruments are limited to nearby sources
because of absorption of the flux on interstellar light. A 
Mediterranean telescope \citep{Antares,Nestor,NEMO} 
and AMANDA/IceCube \citep{AMANDA,ICECUBE}
will survey the whole sky.

\section{Neutrino flux from gamma ray observations}

Let's set out to test the assumption that TeV photons 
produced by a supernova remnant or by an active galaxy 
are of hadronic origin, i.e. that they are the
decay products of $\pi^0$'s. Neutral pions can be produced 
by accelerated protons in either $pp$ or $p\gamma$ interactions 
depending on the relative target density of photons and protons 
in the source region where the protons are accelerated \citep{hooper}. The
$\nu_\mu + \bar\nu_\mu$ neutrino flux ($dN_\nu/dE_\nu$) produced
by the decay of charged pions in the source can be derived from the 
observed $\gamma$-ray flux $(dN_\gamma/dE_\gamma)$ by imposing energy 
conservation \citep{stecker79,salamon96}:
\begin{equation}
\int_{E_{\gamma}^{\rm min}}^{E_{\gamma}^{\rm max}}
E_\gamma {dN_\gamma\over dE_\gamma} dE_\gamma = K
\int_{E_{\nu}^{\rm min}}^{E_{\nu}^{\rm max}} E_\nu {dN_\nu\over dE_\nu} 
dE_\nu~,
\label{conservation}
\end{equation}
where ${E_{\gamma}^{\rm min}}$ ($E_{\gamma}^{\rm max}$) is the minimum
(maximum) energy of the photons that have a hadronic origin.
${E_{\nu}^{\rm min}}$ and ${E_{\nu}^{\rm max}}$ are the corresponding 
minimum and maximum energies of the neutrinos.
The factor $K$ depends on whether the $\pi^0$'s are of 
$pp$ or $p\gamma$ origin. Its value can be obtained from routine particle 
physics. In $pp$ interactions 1/3 of the proton energy goes into each pion 
flavor on average. In the pion-to-muon-to-electron decay chain 2 muon-neutrinos
are produced with energy $E_\pi/4$ for every photon with energy 
$E_\pi/2$ (on average). Therefore the energy in neutrinos matches 
the energy in photons and $K=1$.

Important parameters in relating the neutrino flux to the $\gamma$-ray
flux are the maximum and minimum
energies of the produced photons and neutrinos. 
They appear as limits in the
integrals in Eq.\,(\ref{conservation}). The maximum
neutrino energy is fixed by the maximum energy of the accelerated
protons ($E_p^{\rm max}$) which can be conservatively
obtained from the maximum observed $\gamma$-ray energy $E_{\gamma}^{\rm max}$. 
Following our 
previous assumptions:
\begin{equation}
E_p^{\rm max}=6 E_{\gamma}^{\rm max}~,~~~E_{\nu}^{\rm max}={1\over 12} E_p^{\rm max}~,  
\label{enumax}
\end{equation}
for the $pp$ case. 
The minimum neutrino energy is fixed by the
threshold for pion production. For the $pp$ case:
\begin{equation}
E_p^{\rm min}=\Gamma ~ {(2m_p+m_\pi)^2-2m_p^2 \over 2m_p},
\label{ethpp}
\end{equation}
where $\Gamma$ is the Lorentz factor of the accelerator 
relative to the observer. One has to keep in
mind that in blazar jets the shocked regions where protons can be 
accelerated move relativistically along the jet and are therefore boosted 
relative to the observer. 
The average minimum neutrino energy 
is obtained from $E_p^{\rm min}$ using Eq. (\ref{enumax}).

For $p\gamma$ interactions it can easily be shown,
using the $\Delta$-resonance approximation, that $K=4$. 
The minimum proton energy is given by the 
threshold for production of pions via the
$\Delta$-resonance, and the maximum neutrino 
energy is $5\%$ of the maximum energy to which
protons are accelerated. More detailed calculations
will be published elsewhere \citep{sources}.  

\subsection{Neutrino events from RX~J1713.7-3946}

Recently the CANGAROO collaboration has shown that 
the energy spectrum of the $\gamma$-ray emission from 
the supernova remnant RX J1713.7-3946 matches that expected
if the $\gamma$-rays are the products of $\pi^0$ decay generated
in $pp$ collisions.  
We normalize the neutrino flux 
from this source using Eq.~(\ref{conservation}).
We calculate the total energy emitted   
in $\gamma$-rays  
integrating the $\gamma$-ray spectrum from $\pi^0$ decay
calculated by \citep{CANGAROO-RX}. Gamma rays of 
energies up to $\sim 10$ TeV have been observed, indicating 
that protons are accelerated up to $E_p^{\rm max}\sim 100$ TeV.  
Eq.~(\ref{enumax}) then gives $E_\nu^{\max}\sim 10$ TeV.
$E_\nu^{\min}$ is obtained from Eqs.~(\ref{enumax}) and (\ref{ethpp}), 
where we have taken $\Gamma\sim 1$ assuming the velocity of 
expansion of the supernova shell is not much larger than $\sim$ 10,000 km/s. 
The data in \citep{CANGAROO-RX} has been fitted with a model that
uses a $\sim E^{-2}$ input proton spectrum.
We assume that all charged pions decay in the 
environment of the supernova and hence the neutrino spectrum
follows the input proton spectrum. Under these assumptions
our calculated $\nu_\mu+\bar\nu_\mu$ flux is given by:

\begin{equation}
{dN_\nu\over dE_\nu}=
4.14\times 10^{-11}~(E_\nu/1~{\rm TeV})^{-2}~~~[{\rm cm^{-2}~s^{-1}~TeV^{-1}}].
\end{equation}  

The neutrino event rate is calculated in a straightforward
manner by convoluting the $\nu_\mu+\bar\nu_\mu$ flux with the 
probability of detecting a muon produced in a muon-neutrino 
interaction \citep{Physrep}:
\begin{equation}
N_{\rm events}=A_{\rm eff}\times T \times \int~{dN_\nu\over
dE_\nu}(E_\nu)
P_{\nu\rightarrow\mu}(E_\nu) dE_\nu,
\end{equation}
where $T$ is the observation time and $A_{\rm eff}$ is 
the effective area of the detector. 

\begin{figure}
\plotone{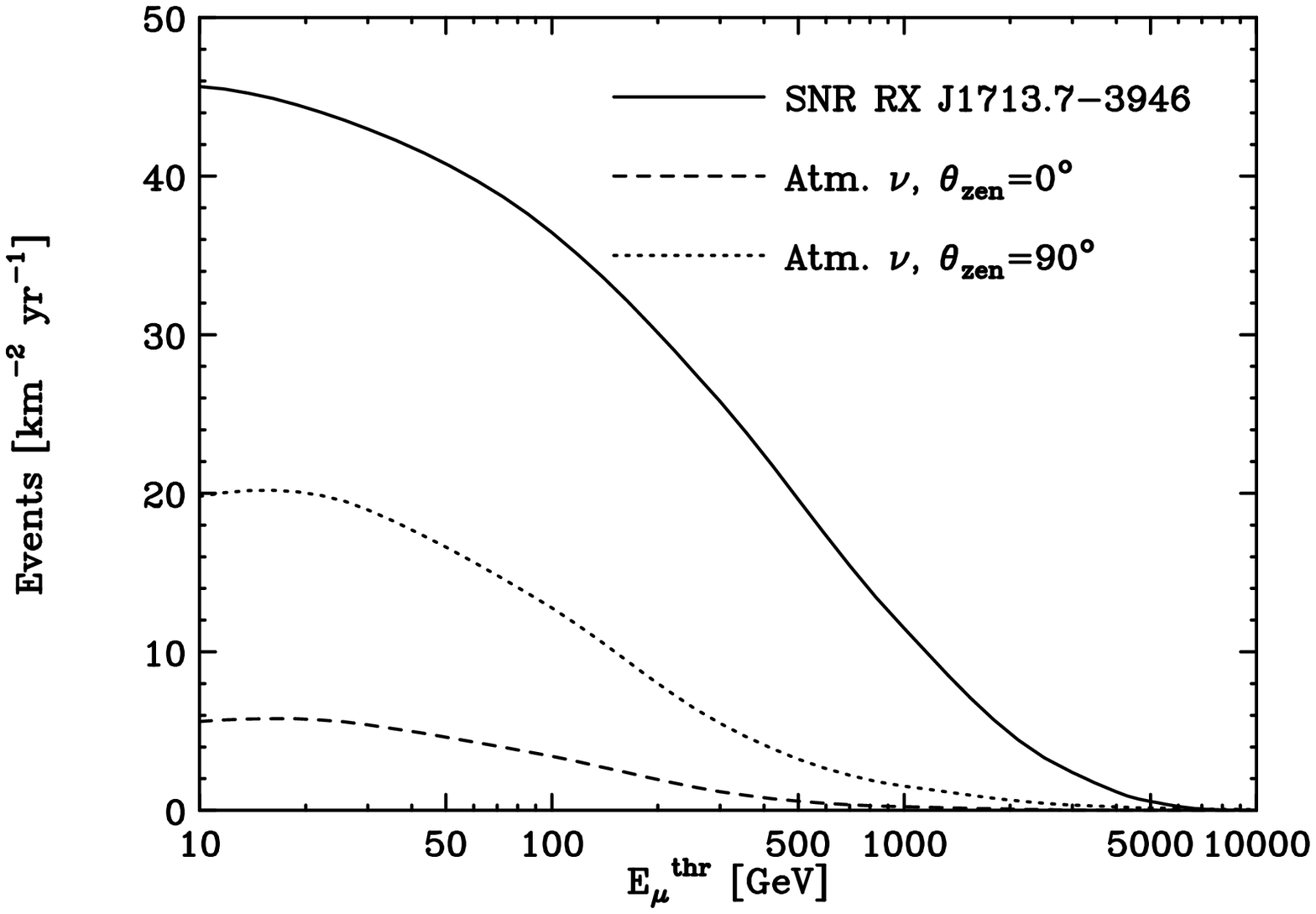}
\caption{Neutrino event rate per year from RX~J1713.7-3946
as a function of the muon
energy threshold in a km-scale neutrino telescope.
Also shown is the event rate expected in a $1^\circ \times 1^\circ$
bin in the sky from atmospheric neutrinos. The dashed curve
shows the prediction for vertical downgoing atmospheric neutrinos and
the dotted the corresponding one for horizontal atmospheric neutrinos.
\label{fig:rate}}
\end{figure}

The neutrino event rate in a detector of effective area 
$A_{\rm eff}=1~{\rm km^2}$ is shown in Fig.~\ref{fig:rate} 
as a function of the muon energy threshold of the detector 
($E_\mu^{\rm thr}$). 
Also shown is the atmospheric neutrino background 
in a $1^\circ \times 1^\circ$ window in the sky.
The dashed and dotted curves show the contribution from vertical 
downgoing and horizontal atmospheric neutrinos respectively, bracketing the 
maximum and minimum values of the expected atmospheric 
neutrino background \citep{volkova}. The actual background 
depends on the (in general) variable zenith angle
of the source as seen from the detector. Clearly  
the background does not pose a problem for the 
identification of this source. 
Due to the much larger muon flux produced in
cosmic ray interactions in the atmosphere compared
to the expected muon flux from neutrino interactions, only neutrinos
going through the Earth can be identified. For this reason
a southern hemisphere source such as RX J1713.7-3946 is
only accesible to a northern hemisphere neutrino telescope. 

In a few years of operation a detector in the northern 
hemisphere such as the current 
Antares ($A_{\rm eff}\sim 0.1~{\rm km^2}$, $E_\mu^{\rm thr}\sim 40$ GeV) 
will be able to identify RX~J1713.7-3946 with a Poisson chance
probability that the event rate is a fluctuation of the atmospheric
neutrino background smaller than $1\%$. 
The IceCube neutrino telescope 
($E_\mu^{\rm thr}\sim 200$ GeV) will do the same job 
in just 1 year of operation if a source
similar to RX~J1713.7-3946 exists in the 
northern hemisphere. Besides, these detectors have the 
potential to discover new SNR's that might
have escaped the scrutinity of $\gamma$-ray Cherenkov 
telescopes. 

It should be mentioned that we have not attempted to correct 
for absorption of the $\gamma$-ray flux in the source or in 
the interstellar medium. The latter is justified since the 
distance to the source is poorly known. For these reasons 
our calculation could well be conservative. We have not 
included the possible effects of oscillation of the 
cosmic beam \citep{crocker}. 

The claims made by the CANGAROO collaboration have been challenged
in a recent paper \citep{Reimer02}. The statement 
is that the EGRET experiment detected less gamma rays from
the associated source 3EG J1714-3857 than expected from
$\pi^0$ decay. However, the EGRET measurements were made 
between 1991 and 1995 \citep{3EGcatalog},
i.e. several years before the CANGAROO observations which
were carried out in 2000-2001. It is possible that 
the source might have experimented large flux variations
over the period of time separating both observations. It 
is not the purpose of our work to further discuss this point.
We stress here that neutrino
telescopes have the sensitivity to settle this important debate.

\subsection{Are there other proton cosmic accelerators?}

Gamma rays at TeV energies have also been detected
from other shell-type SNRs such as 
the Crab Nebula, Cassiopeia A and   
upper limits have been reported from 
the SNR-like shell Sagittarius A East 
at the Galactic center \citep{SgrAWhipple}. Besides,
TeV $\gamma$-rays have been observed from at least 
two blazars: Mkn 421 and Mkn 501. For all these objects the 
observations are less constrained than 
in the case of RX~J1713.7-3946, and in fact 
models in which the $\gamma$-rays are produced by
inverse Compton scattering of ultra-high energy electrons
on ambient photons can be stretched to explain the 
TeV observations. For Cassiopeia A and 
Sagittarius A East we proceed as in the case of 
RX~J1713.7-3946 normalizing the neutrino flux 
by the $\gamma$-ray flux expected from $\pi^0$ decay
calculated in \citep{icrc99CASA} and \citep{meliaSgrA} respectively.
For the Crab we assume that the 
measured high energy $\gamma$-ray spectrum 
has been produced in $pp$ collisions, and the 
corresponding gamma flux from the Mkn 421 and Mkn 501 in 
$p\gamma$ collisions, 
and we compute the accompanying neutrino flux and neutrino event
rate using Eq.~(\ref{conservation}).

We perform the actual calculations by assuming that both 
the photon and the neutrino fluxes are generated by protons with a
differential energy spectrum proportional to $E^{-2}$, and 
assume the neutrino spectrum is parallel to the proton 
spectrum. The spectral index of
photons is taken from the observations assuming the $\gamma$-rays 
are of hadronic origin. 

In the latter calculations we, conservatively, only identify the 
observed 100 GeV-TeV
$\gamma$-rays with $\pi^0$ origin. This spectrum does of
course continue down to low energies and turns over at
$m_{\pi^0}$/2, as it does for RX J1713.7-3946. This low
energy spectrum is not revealed by the observations and
modeling it is not straightforward because of possible
cascading, boost factors, etc. We therefore prefer to
limit the spectral information to the region of observation,
(corresponding to photon energies above a certain value which
we denote by $E_\gamma^{\rm cut}$), thus producing a lower
limit on the neutrino flux. 

High-energy photons can be absorbed in the interstellar infrared
radiation. Because the amount of absorption depends
on the distance to the source and on the poorly known infrared background
\citep{dejager} we are simply going to use as $E_{\gamma}^{\rm max}$ the
maximum
observed energy keeping in mind that this assumption leads to an
underestimate of the neutrino flux.
We have used the measured TeV $\gamma$-ray fluxes given in
the following references: Crab \citep{HEGRACrab},
Mkn 421 \citep{WhippleMkn421c},
and Mkn 501 \citep{HEGRAMkn501-99}.
 
The event rates per ${\rm km^2}$, and the values
of the relevant parameters are shown in Table~\ref{tab:rates}.
We calculated the event rate for two choices of
$E_{\gamma}^{\rm cut}$, namely, the energy threshold of
the Cherenkov telescope that performed the measurement,
and $E_{\gamma}^{\rm cut}=100$ GeV, illustrating the
sensitivity of the neutrino event rate to $E_{\gamma}^{\rm cut}$.
Also tabulated are the background of
atmospheric neutrinos above $E_\nu^{\rm min}$ in
a $1^{\rm o}\times 1^{\rm o}$ pixel in the direction
of the corresponding source, and the Poisson chance
probability of obtaining a background of events
from atmospheric neutrinos equal to the total expected
signal during the time of the observation.

For the SNR Cassiopeia A we obtain  
5 events per ${\rm km^2}$ per year with a $4.7\%$ chance 
probability that the rate is a fluctuation
of the atmospheric background. Due to the extreme inferred energetics
of Sagittarius A East, about 40 times the energy released in a 
typical supernova explosion \citep{meliaSgrA}, the neutrino 
event rate is large:  
$\sim$ 140 ${\rm km^{-2}~yr^{-1}}$.  
Sagittarius A East is however too close to the Galactic center
to be observed by a detector at the South Pole but it could
be detected in a northern hemisphere telescope.  

We conclude that the hadronic origin of the observed TeV-gamma rays 
can be confirmed or disproven with confidence given that neutrino
telescopes will be operated over many years. Because of the total 
transparency of the Universe to neutrinos, we also expect a larger 
number of sources. 
RX~J1713.7-3946 awaits to be detected by northern 
hemisphere telescopes \citep{Antares,NEMO,Baikal,Nestor}. 
Its detection will be a milestone in 
neutrino astronomy. Our results are also encouraging for the
AMANDA/IceCube detectors at the South Pole. They clearly have 
the capability to identify cosmic accelerators observed in 
$\gamma$-rays. They may also reveal sources that have 
escaped detection by Cherenkov telescopes. 

\acknowledgments
We thank Diego F. Torres for discussions. 
We are grateful to the anonymous referee for valuable comments
and suggestions on the manuscript. 
J.A.-M. is supported by NASA Grant NAG5-10919 and thanks the 
Department of Physics, University of Wisconsin, Madison
where part of this work was done. This work was
supported in part by DOE grant No.~DE-FG02-95ER40896 
and in part by the Wisconsin Alumni Research Foundation.

\clearpage

\clearpage

\begin{table}
\renewcommand{\arraystretch}{1.5}
\begin{center}
\caption{
$N_{\nu_\mu+\bar\nu_\mu}$ events per $\rm km^2$
from the Crab and Mkn421 in one year and from Mkn 501
during the 8 months of flaring activity in 1997.
We are assuming that protons of energy up to $10^{20}$ eV can be
accelerated in the Mkn's.
Also shown is the expected rate of
atmospheric neutrino events (assuming Volkova's flux) in a
$1^{\rm o} \times 1^{\rm o}$ bin on the sky
around the position of the source \citep{catalog}
as seen from IceCube. The last row
is the chance probability of observing a background of atmospheric
neutrino events equal to the total signal in the same period of time.
See text for references.
\label{tab:rates}}
\vskip 0.5cm
\begin{tabular}{c|c c|c c|c c}
\tableline\tableline
Source & \multicolumn{2}{c|}{Crab} & \multicolumn{2}{c|}{Mkn 421} & \multicolumn{2}{c}{Mkn 501} \\  \hline
 
$E_{\gamma}^{\rm cut}$ [GeV]& ~100~ & ~500~ & ~100~ & ~260~ & ~100~ & ~500~ \\
 
$E_{\gamma}^{\rm max}$ [TeV]& ~20~ & ~20~ & ~17~ & ~17~ & ~24~ & ~24~ \\
 
$E_{\nu}^{\rm max}$ [TeV]& ~10~ & ~10~ & ~$10^7$~ & ~$10^7$~ & ~$10^7$~ & ~$10^7$~ \\
 
$N_{\nu_\mu+\bar\nu_\mu}$ & ~9.5~ & ~4.0~ & ~7.5~ & ~3.5~ & ~8.0~ & ~5.0~ \\
 
$N_{\nu_\mu+\bar\nu_\mu}^{\rm atm}$ & ~3.5~ & ~3.5~ & ~1.5~ & ~1.5~ & ~1.0~ & ~1.0~ \\
 
Poisson prob. $(\%)$ & $~1.2~$ & $~18.8~$ & ~0.1~ & ~3.8~ & ~0.001~ & ~0.3~ \\
 
\tableline
 
\end{tabular}
 
\end{center}
\end{table}

\end{document}